\documentclass[manuscript]{geophysics}
\usepackage[english]{babel}
\usepackage{amsfonts,epsfig,setspace,bm}
\usepackage{amsmath}
\usepackage{amssymb}
\usepackage{graphicx}
\usepackage{subfigure}
\usepackage{color}

\usepackage[font=doublespacing]{caption}

\newtheorem{theorem}{Theorem}[section]

\usepackage{booktabs}

\usepackage{caption}
\usepackage{lipsum}
\begin{document}

\title{Learning on the correct class for domain inverse problems of gravimetry}
\address{
\footnotemark[1] School of Science, Harbin Institute of Technology, Shenzhen, Shenzhen 518055, China. Email: {\bf liwenbin@hit.edu.cn}
}
\author{Yihang Chen\footnotemark[1] and Wenbin Li\footnotemark[1]}

\footer{}
\lefthead{Chen \& Li}
\righthead{Learning on the correct class}
\date{}
\maketitle

%%%%%%%
\begin{abstract}
%We propose the strategy of learning on the correct class for domain inverse problems of gravimetry. Due to ill-posedness of the inverse gravimetry, the reliability of end-to-end learning approaches is questionable. To deal with this problem, we employ the well-poseness theorems
We consider end-to-end learning approaches for inverse problems of gravimetry. Due to ill-posedness of the inverse gravimetry, the reliability of learning approaches is questionable. To deal with this problem, we propose the strategy of learning on the correct class. The well-posedness theorems are employed when designing the neural-network architecture and constructing the training set. Given the density-contrast function as a priori information, the domain of mass can be uniquely determined under certain constrains, and the domain inverse problem is a correct class of the inverse gravimetry. Under this correct class, we design the neural network for learning by mimicking the level-set formulation for the inverse gravimetry. Numerical examples illustrate that the method is able to recover mass models with non-constant density contrast.
\end{abstract}

%%%%%%%%
\section{Introduction}
The gravimetry approach is of great value in large-scale geophysical explorations.
The inverse problem of gravimetry aims to determine the earth's mass distribution from data of the gravity potential or related gravity quantities \citep{liyaoguo19983}. Because of the equivalent source principle of gravity potential, the inverse problem of gravimetry is severely ill-posed in the Hadamard sense \citep{isa90}. To regain well-posedness, some a priori conditions should be imposed on the anomalous mass distribution, restricting the solution into a correct class. Based on the uniqueness theorems \citep{isa90}, Isakov and his collaborators developed a level-set framework to solve domain inverse problems of gravimetry \citep{isaleuqia11,luleuqia15,liluqia16,liqia21,liqia22}, which is one correct class of inverse gravimetry problems. Consider the volume mass distribution $\mu=f\chi_D$, where $\chi_D$ denotes the characteristic function of a domain $D$, and $f$ is a density-contrast function. Given some a priori information of $f$ and imposing certain geometric constraints on $D$, the domain inverse problem of gravimetry admits a unique solution. In the level-set approach, a level-set function $\phi$ is employed to describe the unknown domain $D$, and an iterative algorithm is developed to solve the level-set function by fitting the measurement data.

In recent years, deep learning approaches are widely used to solve inverse problems in many fields of science. Different from classical iterative algorithms, an end-to-end learning method aims to develop a neural-network architecture which can generate recovered solutions directly by inputting measurement data. The learning approaches have the potential to handle high-level noise contamination of data; moreover, the end-to-end learning methods are much more efficient than classical iterative algorithms. Due to these advantages, using deep neural networks and learning strategies has arisen as a new trend in the study of inverse gravimetry \citep{hualiuqizha21,yanhuliujiewanche22,zhazhaliufan22,zhazhafanma22,zhochelvwan23}. On the other hand, the learning approaches have limitations in several aspects, including the reliability of solutions and the generalization ability of networks. Especially in the inverse gravimetry, due to the severe ill-posedness of the inverse problem, the end-to-end neural networks may have poor performance when moving to a test set deviating from the training set. In this work, we try to deal with this problem in domain inverse problems of gravimetry. Different from the popular trend that promotes enlarging the training set and employing diverse models \citep{hualiuqizha21,zhochelvwan23}, we propose the strategy of learning on the correct class.

%%%%%%%
\section{Learning on the correct class}
\label{Sec2}
The ill-posed inverse problems can be solved only on a conditionally correct class. For the inverse gravimetry problem, since the solution is theoretically nonunique, there may exist infinitely many structures that reproduce a given set of measurement data within the same level of accuracy. Even though the learning approaches have the magic to produce reasonable solutions at times, the reliability of solutions is questionable if not restricted on the correct class. 
%In this context, considering diverse models or data augmentation \cite{} is probably not the key to achieving reliable generalization ability for the learning approaches.

Let $U$ denote the gravity potential generated by a mass distribution $\mu$ with $\mathrm{supp}\mu\subset\Omega$,
\begin{equation}\label{eqn1}
U(\mathbf{y};\mu)=\gamma\int_{\Omega} K(\mathbf{y},\mathbf{x})\mathrm{d}\mu(\mathbf{x}),
\end{equation}
where $K(\mathbf{y},\mathbf{x})=K(|\mathbf{y}-\mathbf{x}|)$ is the fundamental solution of Laplace's equation,
\begin{equation}\label{eqn2}
K(\mathbf{y},\mathbf{x})=\left\{\begin{array}{ccc}
-\frac{1}{2\pi}\mathrm{ln}|\mathbf{y}-\mathbf{x}| &,&\mathbf{x},\mathbf{y}\in\mathbf{R}^2\,,\\
\frac{1}{4\pi|\mathbf{y}-\mathbf{x}|} &,&\mathbf{x},\mathbf{y}\in\mathbf{R}^3\,,
\end{array}\right.
\end{equation}
and $\gamma$ is a constant related to the universal gravitational constant. The vector $\nabla U(\mathbf{y};\mu)$ represents the gravity force at $\mathbf{y}$. The inverse problem of gravimetry reads as follows: Given the gravity force $\nabla U(\mathbf{y};\mu)$ on $\Sigma_0\subset\mathbf{R}^n\setminus\Omega$, find the mass distribution $\mu$ with $\mathrm{supp}\,\mu\subset\Omega$.

In this work, we consider a conditionally correct class of the inverse problem as the volume mass distribution takes the form of $\mu=f\chi_D$, where $f(\mathbf{x})$ is a density-contrast function, and $\chi_D$ denotes the characteristic function of the domain $D$: $\chi_D(\mathbf{x})=1$, $\mathbf{x}\in D$; $\chi_D(\mathbf{x})=0$, $\mathbf{x}\notin D$. In this formulation, we have the following uniqueness results for the inverse problem of gravimetry \citep{isa90,liqia21}.
\begin{theorem}\label{thm1}
Let $\Omega_0$ be a convex domain with analytic (regular) boundary, $\Sigma_0\subset\partial\Omega_0$  be a nonempty hyper-surface, and $\Omega\subset\Omega_0$ be a bounded domain with connected $\mathbf{R}^n\setminus\overline{\Omega}$. $D\subset\Omega$ denotes the domain of mass anomaly having piecewise smooth boundaries. Given the gravity force $\nabla U$ on $\Sigma_0$ and given $f\ge0$ in $\Omega$, the domain $D$ can be uniquely determined if one of the following constraints is satisfied:

(1) $D$ is star-shaped with respect to its center of gravity, and $f$ is constant;

(2) $D$ is convex in one direction, e.g. in $x_n$, where $x_n$ denotes a component of the spatial coordinate $\mathbf{x}=(x_1,\cdots,x_n)\in\mathbf{R}^n$, and $f$ is constant;

(3) $D$ is convex in $x_n$, $f$ does not depend on $x_n$, $f\in C(\Omega)$, and $\Omega\subset \mathrm{supp}\,f$;

(4) $D$ is convex, $f\in L_1(\Omega)$, and $0<f$ on $\Omega$.
\end{theorem}

Theorem \ref{thm1} indicates that the domain of mass can be uniquely determined if the density-contrast function is given and certain constraints are imposed. The domain inverse problem is a correct class of the inverse gravimetry. Under the correct class, the learning approaches are more likely to achieve reliable solutions. 

In the level-set approach \citep{isaleuqia11,liqia21}, a level-set formulation is proposed for the volume mass distribution $\mu=f\chi_D$,
\begin{equation}\label{eqn3}
\mu(x)=f(x) H(\phi(x))\,,
\end{equation}
where $\phi(x)$ is the level-set function and $H(\cdot)$ is the standard Heaviside function,
\begin{eqnarray}
\phi(x)\left\{\begin{array}{ccc}\ge 0&,&x\in\bar{D}\\ <0&,&x\in\bar{D}^C \end{array} \right., \quad H(\cdot)=\left\{\begin{array}{ccc}1&,&x\ge0\\ 0&,&x<0 \end{array} \right.\,.
\end{eqnarray}
Mimic to this expression, we propose the following formula for $\mu$ in the learning approach,
\begin{equation}\label{eqn4}
\mu(\cdot)=f\, \sigma\left(\tilde{\Lambda}_\Theta(\cdot) \right)\,,
\end{equation}
where $\tilde{\Lambda}_\Theta(\cdot)$ denotes a neural network with the parameter $\Theta$, and $\sigma(\cdot)$ is the sigmoid activation function,
\begin{equation}
\sigma(s)=1/(1+e^{-s})\,,
\end{equation}
which can be viewed as a smooth version of the Heaviside.

Researchers in geophysics have the intuition to consider the so-called sparse inversion in learning approaches for inverse gravimetry \citep{hualiuqizha21}, which is essentially the domain inverse problem. While they suggest diverse models or data augmentation to improve the network generalization ability, we consider an opposite direction to achieve reliable solutions, i.e. restricting the solution on the correct class according to uniqueness theorems.

%%%%%%%%%%%%
\section{Network architecture and training}
\subsection{Building training set on the correct class} \label{Sec3.1}
Consider a given geometry of exploration: $\Omega=(0,1)\times(0,0.5)$\,km, and $\Sigma_0=\{(x_1,x_2)\mid 0\le x_1\le 1, x_2=-0.01\}$\,km; $256$ measurements are uniformly sampled along $\Sigma_0$. We build the pairs of mass models and gravity data for the training of neural networks.

According to Theorem \ref{thm1}, we construct the mass model $\mu$ on a correct class of the domain inverse problem. The constraint of star-shape or convexity seems too strong for the unknown domain $D$, which avoids the condition of disconnected regions. But the convexity in one direction is reasonable in many applications. As a result, we focus on situations (2) and (3) in Theorem \ref{thm1}, where (2) can be viewed as a special case of (3).

\begin{figure}
\centering
(a){{\includegraphics[scale=0.56,angle=0]{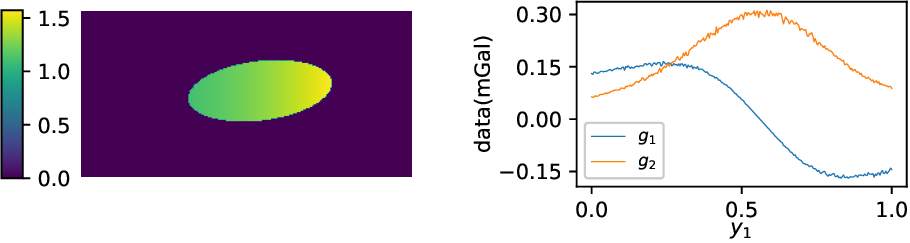}}}
(b){{\includegraphics[scale=0.56,angle=0]{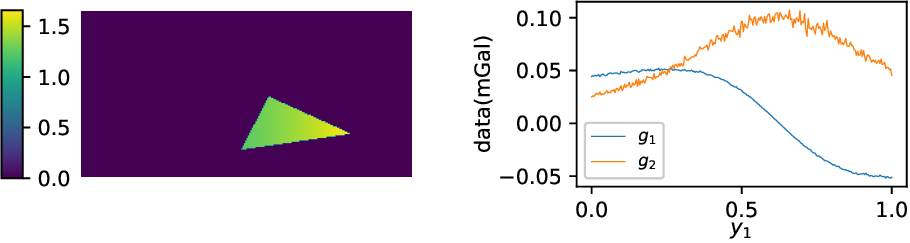}}}
(c){{\includegraphics[scale=0.56,angle=0]{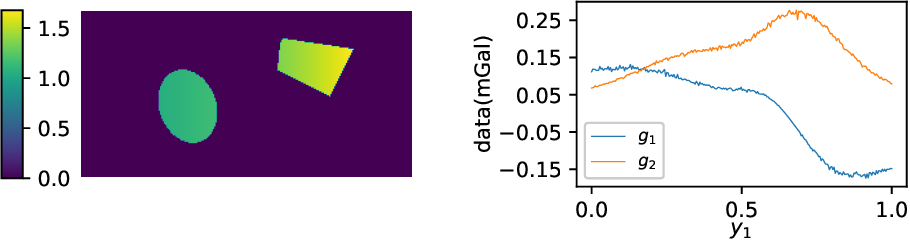}}}
(d){{\includegraphics[scale=0.56,angle=0]{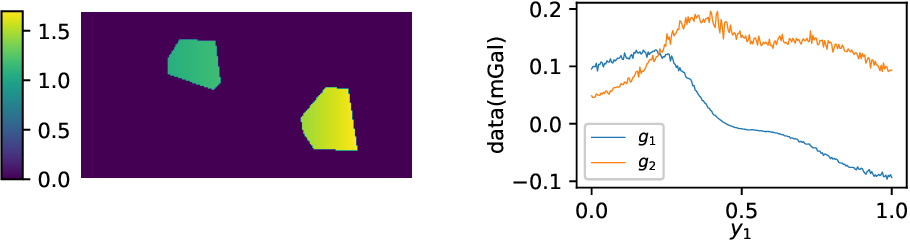}}}
\caption{Training set on the correct class. (a)-(d) show 4 examples of the 20,000 models we have built. The left column plots the mass distributions $\mu$, and the right column plots the simulated measurement data $\nabla U$ with $0-5\%$ Gaussian noises.}
\label{Fig1}
\end{figure}

The left column of Figure \ref{Fig1} shows 4 examples of 20,000 mass models we have built. The computational domain $\Omega$ is divided into $256\times128$ grids. The mass model $\mu$ is composed of ellipses and polygons, which are both convex, and the overall mass model is convex in $x_2$, i.e. the vertical direction. The density-contrast function $f$ is given as a priori information; e.g. $f(x_1,x_2)=1+x_1^2$ in this application, which satisfies the constraint (3) of Theorem \ref{thm1}.
The gravity data $\nabla U$ is calculated on $\Sigma_0$ according to the forward model (\ref{eqn1}). $0-5\%$ Gaussian noises are added to $\nabla U$, simulating the situation of practical measurements. The right column of Figure \ref{Fig1} plots the simulated measurement data of the 4 mass models in the training set.

\subsection{Network architecture}
The end-to-end neural network, $\Lambda_\Theta:\ \nabla U\mapsto \mu$, aims to recover the mass model $\mu$ by inputting the gravity data $\nabla U$. As shown in the preceding section, learning on the correct class suggests us to consider the formulation of (\ref{eqn4}):
\[
\Lambda_\Theta(\cdot)=f\, \sigma\big(\tilde{\Lambda}_\Theta(\cdot) \big)\,, 
\]
where $f$ is the given density-contrast function, $\sigma(\cdot)$ is the sigmoid activation function mimicking the Heaviside function, and $\tilde{\Lambda}_\Theta(\cdot)$ is a neural network.

\begin{figure}
\centering
{{\includegraphics[scale=0.6,angle=0]{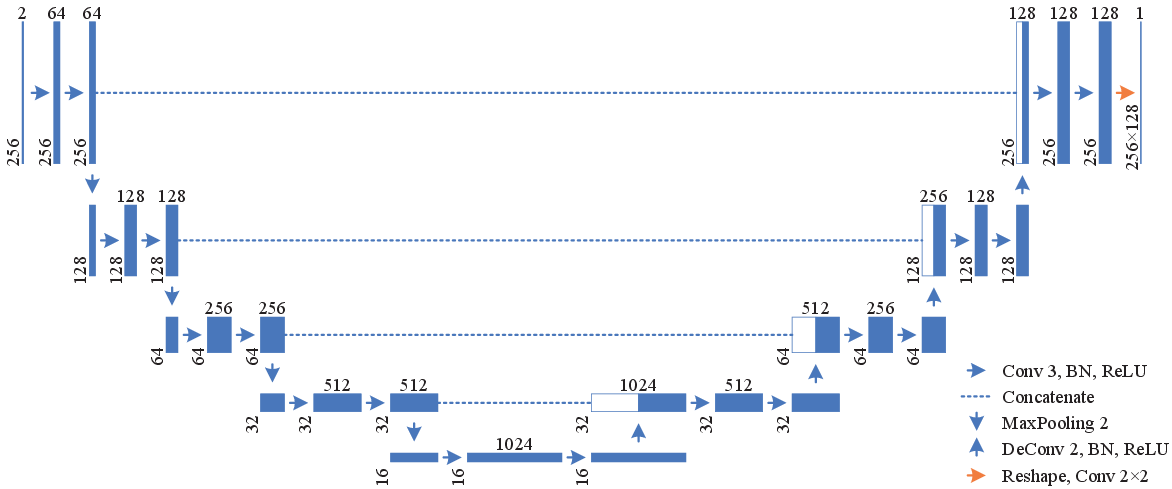}}}
\caption{A modified U-net architecture for $\tilde{\Lambda}_\Theta(\cdot)$. The overall neural network for training is $\Lambda_\Theta(\cdot)=f\, \sigma\big(\tilde{\Lambda}_\Theta(\cdot) \big)$.}
\label{Fig2}
\end{figure}

We propose to use a modified U-net architecture \citep{ronneberger2015u} for $\tilde{\Lambda}_\Theta(\cdot)$, and Figure \ref{Fig2} shows the structure. It is a U-shaped convolutional neural network with 2 channels of input and 1 channel of output, where the 2 channels of input corresponds to the two components of $\nabla U$. In most applications, the measurement surface $\Gamma_0$ is one-dimensional lower than the exploration domain $\Omega$. To map from the data surface $\Gamma_0$ to the model domain $\Omega$, the convolutional neural network $\tilde{\Lambda}_\Theta(\cdot)$ must have the ability to expand one dimension, e.g. $\tilde{\Lambda}_\Theta:\ \mathbf{R}^{n-1}\times\mathbf{R}^{n-1}\to\mathbf{R}^n$. We employ the strategy of joining multiple channels to achieve the dimension extension \citep{yanhuliujiewanche22}. As shown in Figure \ref{Fig2}, the layer before the last has $128$ channels, each of which corresponds to a horizontal slice of the model domain $\Omega$, and the number of channels expands along the $x_2$-direction. The last layer reshapes the output, where a convolution is added after joining the channels. For rigorous math modeling of U-net architectures and general neural networks, we refer readers to \cite{bialaili23}.

\subsection{Training and optimization}
Let $\{(\nabla U_s, \mu_s) \mid s=1,\cdots M\}$ denote the training set built on the correct class. The loss function for training is defined as follows,
\begin{equation}\label{eqn_Loss}
\mathcal{L}_t(\Theta)= \frac{1}{|\mathcal{I}_t|}\sum_{s\in\mathcal{I}_t} \frac{1}{N}\left(\left\|f\,\sigma\big(\tilde{\Lambda}_\Theta(\nabla U_s) \big)  -\mu_s\right\|_2^2+\lambda_1\left\| \sigma\big(\tilde{\Lambda}_\Theta(\nabla U_s) \right\|_{\mathrm{TV}}\right)+\lambda_2\|\Theta_w\|_2^2\,,
\end{equation}
where $\mathcal{I}_t\subset\{1,\cdots M\}$ denotes a mini-batch of the index set, $|\mathcal{I}_t|$ denotes the batch size, and $N$ is the pixel of the mass model, e.g. $N=256\times128$. In (\ref{eqn_Loss}), the 1st term measures the discrepancy between the true model $\mu_s$ and the neural-network output. The 2nd term is a regularization aiming to enhance the piecewise constant property of the recovered domain; since the end-to-end learning approach generates solutions directly, the regularization is introduced in the training process. The TV-norm is implemented as: 
\[
\|\mathbf{p}\|_{TV}=\sum_{i_1,\cdots,i_n}\big(\sqrt{(\mathbf{p}_{i_1+1,\cdots, i_n}-\mathbf{p}_{i_1,\cdots, i_n})^2+\epsilon}+\cdots+\sqrt{(\mathbf{p}_{i_1,\cdots, i_n+1}-\mathbf{p}_{i_1,\cdots, i_n})^2+\epsilon}\,\big), 
\]
where $\epsilon$ is a small constant, e.g. $\epsilon=10^{-10}$. The 3rd term of (\ref{eqn_Loss}) is a standard regularization for the neural network, where $\Theta_w\subset\Theta$ denotes the parameter of convolution kernels. $\lambda_1$ and $\lambda_2$ are two constants controlling the amount of regularization. Similar to \cite{bialaili23}, we use a random shuffling strategy for the sampling of the mini-batch set $\mathcal{I}_t$, and the training parameter $\Theta$ is updated according to the Adam algorithm.

%%%%%%%%%%%%
\section{Results}
\begin{figure}
\centering
(a){{\includegraphics[scale=0.56,angle=0]{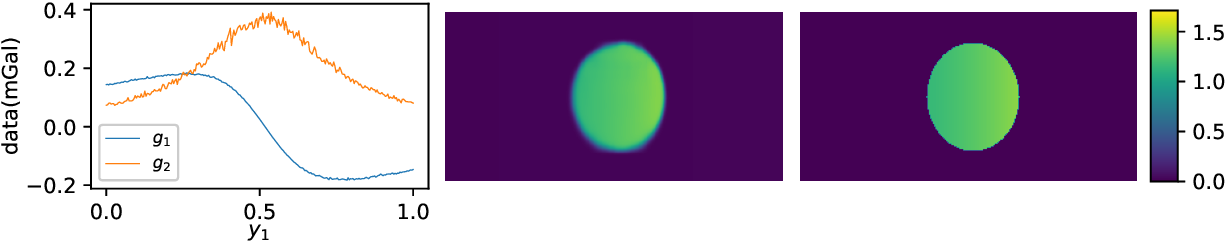}}}
(b){{\includegraphics[scale=0.56,angle=0]{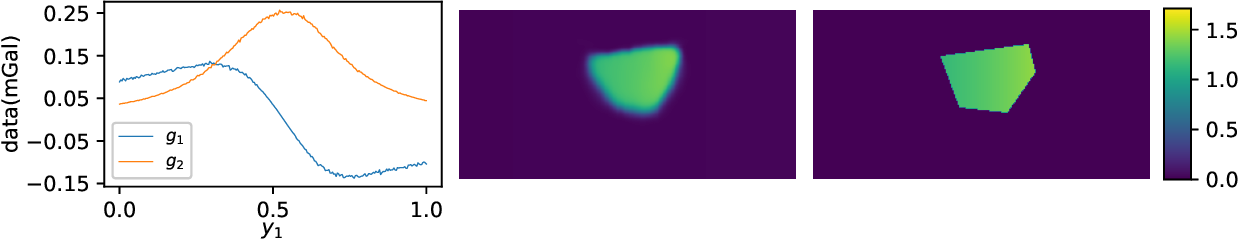}}}
(c){{\includegraphics[scale=0.56,angle=0]{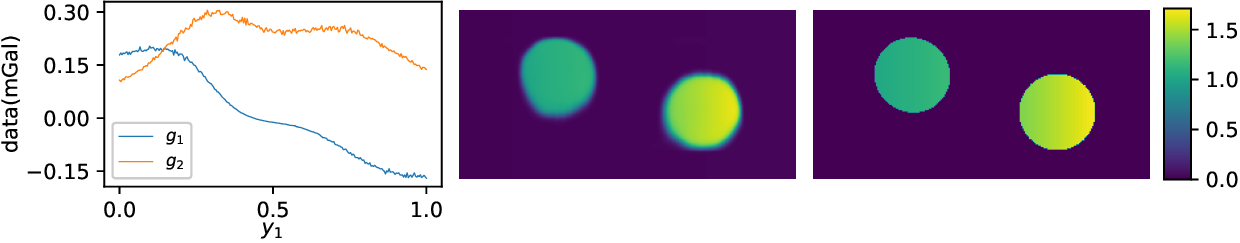}}}
(d){{\includegraphics[scale=0.56,angle=0]{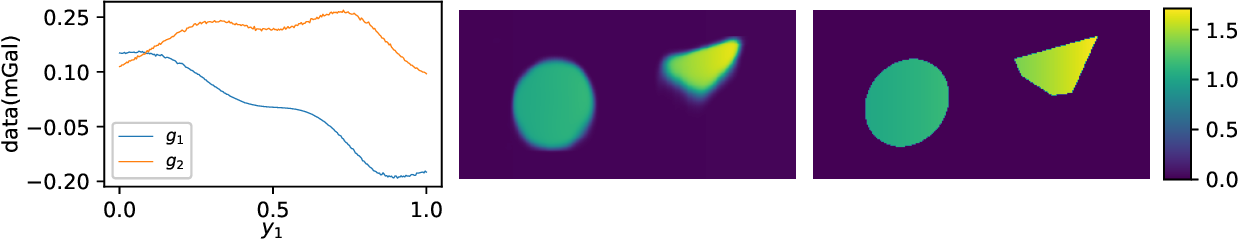}}}
\caption{Results 1: The trained neural network $\Lambda_\Theta$ is implemented on the test set of 2000 samples. (a)-(d) show 4 examples of the 2000 test samples. The 1st column plots the measurement data with $0-5\%$ Gaussian noises, which are the inputs of $\Lambda_\Theta$; the 2nd column plots the recovered solutions; the 3rd column plots the true models.}
\label{Fig3}
\end{figure}

As shown in the preceding section, we have built 20,000 pairs of mass models and gravity data on the correct class. We separate them into 3 sets: the training set of 16,000 pairs, $\{(\nabla U_s, \mu_s) \mid s=1,\cdots 16000\}$, the validation set of 2000 pairs, $\{(\nabla U_s, \mu_s) \mid s=16001,\cdots 18000\}$, and the test set of the remaining 2000 pairs, $\{(\nabla U_s, \mu_s) \mid s=18001,\cdots 20000\}$. The neural network $\Lambda_\Theta$ is trained on the 1st set and validated on the 2nd set, where we take $\lambda_1=10^{-2}$, $\lambda_2=10^{-4}$, and the batch size $|\mathcal{I}_t|=40$.
The trained neural network $\Lambda_\Theta$ is then implemented on the test set of 2000 samples. Figure \ref{Fig3} shows the recovered solutions in 4 of the 2000 samples. Note that the test set has no intersection with the training set. To quantitatively evaluate the performance of reconstructions, we compute the peak signal-to-noise ratio (PSNR) and the structural similarity index (SSIM). Table \ref{Tab1} lists the average values of PSNR and SSIM for the reconstructions on the 2000 tests samples.

\begin{figure}
\centering
(a){{\includegraphics[scale=0.56,angle=0]{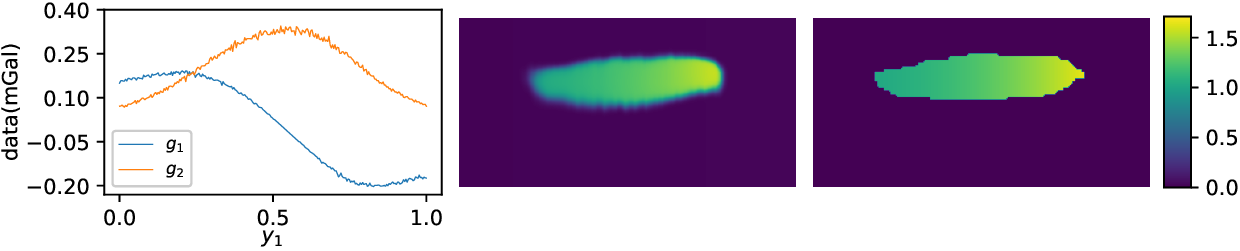}}}
(b){{\includegraphics[scale=0.56,angle=0]{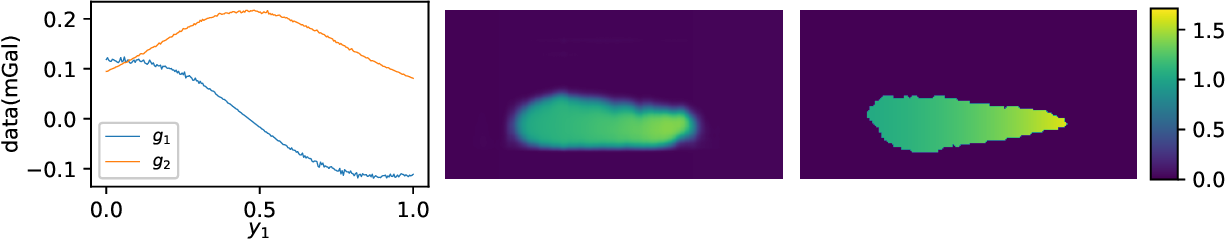}}}
(c){{\includegraphics[scale=0.56,angle=0]{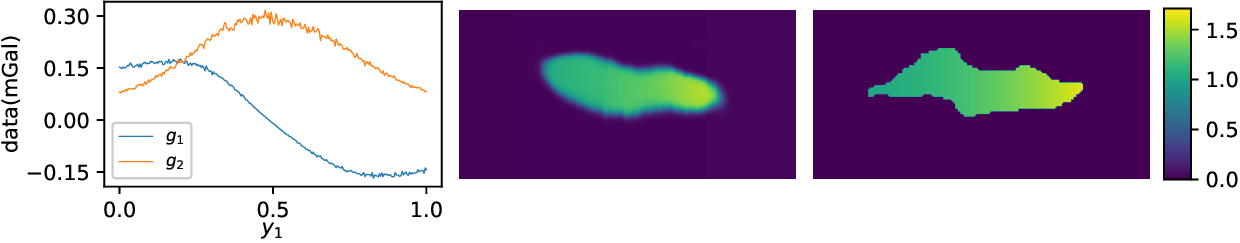}}}
(d){{\includegraphics[scale=0.56,angle=0]{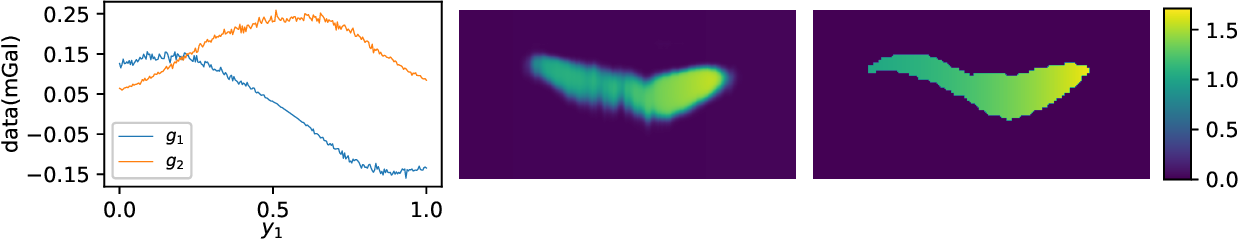}}}
\caption{Results 2: To test the generalization ability, the trained neural network $\Lambda_\Theta$ is implemented on the set of 200 salt models. (a)-(d) show 4 examples of the 200 salt models. The 1st column plots the measurement data with $0-5\%$ Gaussian noises, which are the inputs of  $\Lambda_\Theta$; the 2nd column plots the recovered solutions; the 3rd column plots the true models.}
\label{Fig4}
\end{figure}

To test the generalization ability, we further implement the trained neural network $\Lambda_\Theta$ on a set of 200 salt domes, which has mass models different from those of the training set. The domain of each salt dome is convex in $x_2$, and the density-contrast function is still given as $f=1+x_1^2$, satisfying the uniqueness theorem of inverse gravimetry. Figure \ref{Fig4} shows the results in 4 examples of the 200 samples. Table \ref{Tab1} lists the average values of PSNR and SSIM for the reconstructions on the 200 salt models. Considering the ill-posedness of the inverse problem of gravimetry, the reconstructions are adequate and the learning method is promising. Moreover, the end-to-end neural network is super efficient.

\begin{table}[h]
		\centering
		\begin{tabular}{ccc} 
			\toprule
			\hspace{4pt}           &         on the test set of 2000 samples       &             on the 200 salt models                                    \\ 
			\midrule
			average PSNR      \hspace{4pt}        &        22.64         &            18.62                               \\
			average SSIM      \hspace{4pt}        &         0.91          &              0.79                             \\
			\bottomrule
		\end{tabular}
		
		\caption{List of PSNR and SSIM for the reconstruction results.}
		\label{Tab1}
\end{table}

%%%%%%%%%%%%
\section{Conclusion}
Using learning approaches has arisen as a new trend in the study of inverse gravimetry, but the reliability of learning approaches is questionable. In a mathematical view, the ill-posed inverse problems can be solved only on a conditionally correct class, which should be the same situation when employing the learning approaches. We propose the strategy of learning on the correct class for the domain inverse problems of gravimetry. With the density-contrast function given as a-priori information, the end-to-end neural network is developed to recover the mass model. The well-posedness theorems are employed when designing the neural-network architecture and constructing the training set. Numerical examples demonstrate the efficacy of the proposed method. In geophysical explorations using gravity data, the correct class of solutions should be firstly prescribed, and the proposed method is efficient and promising.

%%%%%%%%%%%%
\section*{Acknowledgments}
Wenbin Li is supported by Natural Science Foundation of Shenzhen (grant no. JCYJ20190806144005645).

\newpage
\bibliographystyle{seg}
\bibliography{myref}

\newpage
%\listoffigures
%\newpage

\end{document}